m.koratzinos@cern.ch

# Note

# Performance limitations of circular colliders: head-on collisions

*M. Koratzinos*
University of Geneva, Switzerland

## Abstract

We review the different performance limitations of circular colliders, namely the limitations of energy reach, maximum attainable luminosity and beam lifetime. This paper considers only head-on collisions. We consider the range of beam energies from 45GeV to 250 GeV and collider circumferences from 20 to 100kms.

# Contents

# 1. Introduction

Recent interest in circular $e^+e^-$ colliders has prompted a debate about ultimate limits of the current technology (and ultimately what it would take to overcome them). We attempt a simplification of the large parameter space with the aim to produce an easy to calculate "hand-out" for circular collider performance.

The major limitations of circular colliders considered here are

- Power consumption limitations that affect the luminosity
- Machine dimension limitations that affect the luminosity and the energy reach
- Beam-beam interaction limitations that affect the luminosity, and finally
- Beamstrahlung (BS) limitations that affect beam lifetimes (and ultimately luminosity)

# 2. Energy reach

In a circular collider the energy reach is a very steep function of the bending radius. The energy loss per turn is given by:

$$E_{loss}[GeV] = 8.85 \times 10^{-5} \frac{E^4}{r_{bend}} \qquad (1)$$

Where $E$ is the beam energy and $r_{bend}$ the bending radius. We consider that a reasonable way to approach this problem is by looking at what percentage of a tunnel of a circular collider would be taken up by the RF acceleration system. Clearly, a circular collider with an RF system longer than the circumference of the machine would make no sense. To get a realistic and quantitative plot, one needs to use assumptions about reasonable accelerating gradients that are achievable and reasonable dipole fill factors. We have used the following assumptions:

- RF gradient: 20MV/m
- Dipole fill factor in the arcs: 85% (for comparison, LEP was 87%)
- RF head room (available voltage over energy loss): 30%

Then the energy reach for a specific ratio of RF system length to the total length of the arcs can be plotted, as shown in Figure 1.

LEP2 had a ratio of RF to bend length of 2.2%. FCC-ee operating at 175GeV and with a bending radius of 11000m [1] would be comfortably sitting below the 1% line. The energy reach of a 11000m radius accelerator can be 236GeV per beam (472GeV $E_{CM}$) for an RF system of 2% the length of the bending system under the assumptions stated above.

# 3. Luminosity

In a circular accelerator and for head-on collisions, luminosity is given by

$$\mathcal{L} = \frac{f_{rev} n_b N_b^2}{4\pi \sigma_x \sigma_y} R_{hg} \qquad (2)$$

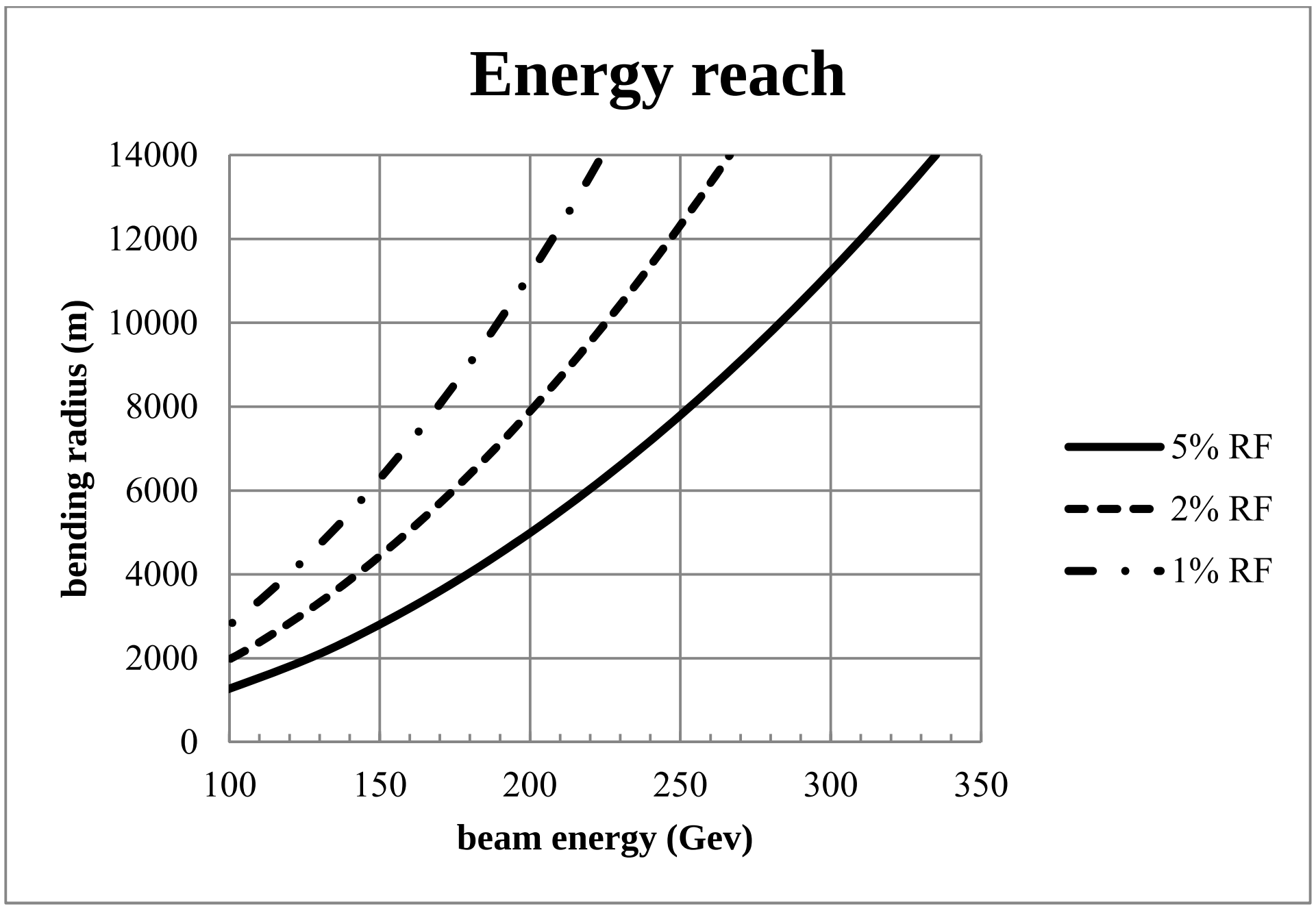

**Figure 1:** Energy reach (in beam energy) of a circular accelerator with different bending radii under three different assumptions of the size of the RF acceleration length compared to the total length of the arcs ($\boldsymbol{2\pi r_{machine}}$ where $\boldsymbol{r_{machine} = r_b/0.85}$).

where $f_{rev}$ is the revolution frequency of the accelerator, $n_b$ is the number of bunches in the machine, $N_b$ is the number of electrons (positrons) per bunch, $\sigma_x, \sigma_y$ is the beam size in x and y, and $R_{hg}$ is the geometric hourglass factor. This can be re-written in terms of the vertical beam-beam parameter, $\xi_y$,

$$\xi_y = \frac{N_b r_e \beta_y^*}{2\pi\gamma\sigma_x\sigma_y} \tag{3}$$

where $r_e = \frac{e^2}{m_e c^2}$ is the classical radius of the electron, $m_e$ its mass, $e$ its charge, c the speed of light, $\beta_y^*$ the vertical beta function at the interaction point, and the total SR power dissipated by one beam $P_{tot}$ ,

$$P_{tot} = \frac{4\pi}{3}\frac{r_e}{{m_e}^3}E^4\frac{f_{rev}n_b N_b}{\rho} \tag{4}$$

where $\rho$ is the bending radius of the machine and $E$ the beam energy. The luminosity formula we can thus derive is

$$\mathcal{L} = \frac{3}{8\pi}\frac{e^4}{{r_e}^4}P_{tot}\ \frac{\rho}{{E_0}^3}\xi_y\frac{R_{hg}}{\beta_y^*} \tag{5}$$

This is a rather simple formula and therefore the maximum luminosity of a circular collider can easily be inferred by knowing the total power consumed, the bending radius of the ring, the maximum vertical beam-beam parameter, the hourglass factor and the vertical beta function. In more convenient units, the formula becomes:

$$\mathcal{L} = 6.0 \times 10^{34}\left(\frac{P_{tot}}{50MW}\right)\left(\frac{\rho}{10km}\right)\left(\frac{120GeV}{E_0}\right)^3\left(\frac{\xi_y}{0.1}\right)\left(\frac{R_{hg}}{0.83}\right)\left(\frac{1mm}{\beta_y^*}\right)cm^{-2}s^{-1} \tag{6}$$

Where the value of $R_{hg} = 0.83$ corresponds to a $\beta_y^*$ value of 1 mm and a longitudinal beam size at the IP of 1.2 mm.

We need to stress that the luminosity in formula (6) is valid only in the case that the accelerator is realizable: one for instance cannot insist on an arbitrary low value of $\beta_y^*$ . However, we believe that the numbers given in parenthesis are numbers that are realisable.

Therefore at a given energy the luminosity of a circular collider is proportional to the SR power dissipated.

The relationship with the bending radius is not as straightforward, as there is a dependence of the maximum achievable $\xi_y$ on the damping decrement which increases with smaller bending radius. For a fixed $\xi_y$ the luminosity increases linearly with the bending radius $\rho$. $\xi_y$ cannot increase beyond a limit usually referred to as the beam-beam limit, ${\xi_y}^{max}$ . If we assume as suggested in [2] that ${\xi_y}^{max} \propto \left({}^{1}/_{\rho}\right)^{0.3\ to\ 0.4}$ then $\mathcal{L}^{max} \propto \rho^{0.6\ to\ 0.7}$ . We need to stress that there is considerable debate regarding the dependence of ${\xi_y}^{max}$ on the damping decrement (and, therefore, $\rho$).

The relationship of luminosity with beam energy is not straightforward either, as at high energies it is increasingly difficult to be able to run at the beam-beam limit without having beam lifetime issues. If the machine is running in the beam-beam dominated regime (see section 5 below), then ${\xi_y}^{max} \propto \left({E_0}^3\right)^{0.3\ to\ 0.4}$, therefore $\mathcal{L}^{max} \propto {E_0}^{-1.8\ to -2.1}$, i.e it approximately drops as the square of the beam energy. For the specific design parameters of FCC-ee the beam-beam regime ends at approximately 170GeV beam energy (see Figure 6). We should also stress that the above formula assumes a constant hourglass factor, which is a simplification (the longitudinal beam size is larger at lower energies and at high BS effects, see section 4 below).

The beam-beam parameter in the horizontal plane is defined in a similar manner to eqn. (3):

$$\xi_x = \frac{N_b r_e \beta_x^*}{2\pi\gamma{\sigma_x}^2} \tag{7}$$

Therefore, imposing that the machine has the same beam-beam parameter in both planes gives the relationship:

$$\xi_x = \xi_y \quad \Rightarrow \quad \frac{\beta_x^*}{\beta_y^*} = \frac{\epsilon_x}{\epsilon_y} \qquad (8)$$

where $\epsilon_x$ and $\epsilon_y$ are the horizontal and vertical emittances of the machine.

To maximise the luminosity for a given machine with a given bending radius and power consumption, one needs therefore to operate at the beam-beam limit (given by machine parameters), decrease $\beta_y^*$ as much as achievable and at the same time keep $R_{hg}$ as close to 1 as possible by ensuring that the longitudinal beam size is kept as small compared to $\beta_y^*$ as possible. To demonstrate the achievable luminosities of circular colliders we have assumed the latest TLEP parameters of $P_{tot} = 50MW$ per beam, $\rho = 11km$, $\beta_y^* = 1mm$, $R_{hg} = 0.75$ (corresponding to a longitudinal bunch length of 2mm), and various values of ${\xi_y}^{max}$. The results can be seen in Figure 2. We need to stress however that such a machine, to be realizable in practice, should yield reasonable beam lifetimes due to beamstrahlung, as discussed in section 4 below.

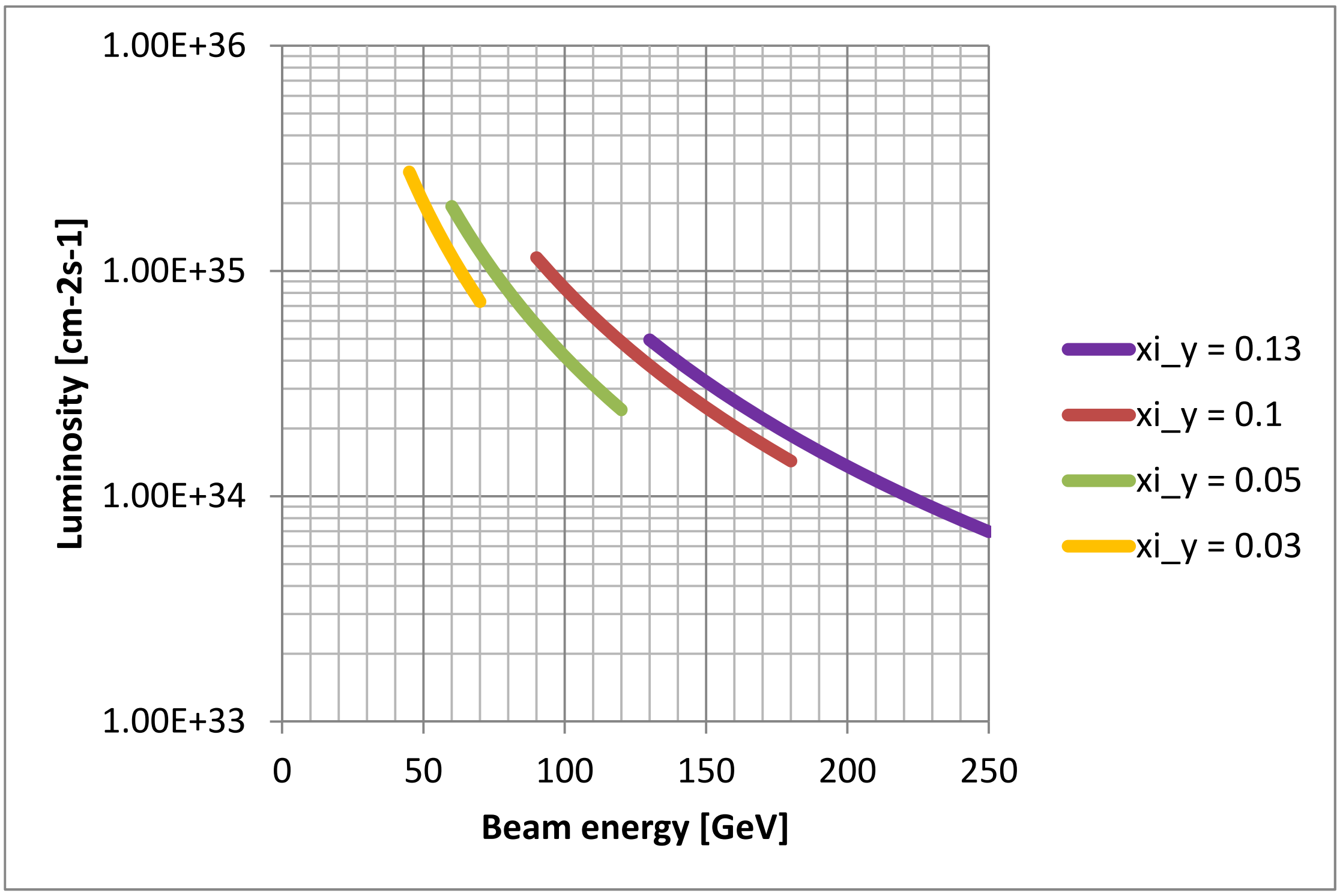

**Figure 2:** Maximum achievable luminosity as a function of beam energy for a circular collider of bending radius 11km, power consumption 50 MW per beam, vertical beta function at the IP of 1mm and longitudinal beam size of 2mm. Different curves are shown corresponding to different maximum vertical beam-beam parameter values.

### 3.1. The hourglass reduction factor

The geometric reduction factor in luminosity when the longitudinal beam size is similar to the focusing of the beams at the interaction point is referred to as the hourglass factor $R_{hg}$. It is not desirable to operate a machine with longitudinal beam sizes much smaller than $\beta_y^*$. Apart from losing luminosity due to geometry, synchrotron-betatron resonances can blow up the beam emittance and lower the luminosity even further.

We have calculated this factor using the approximation in [3]. The variation can be seen in Figure 3. For typical designs as in [1], where the longitudinal RMS bunch length ranges from 120% to 260% of the vertical $\beta^*$ value, it varies between 0.64 and 0.82.

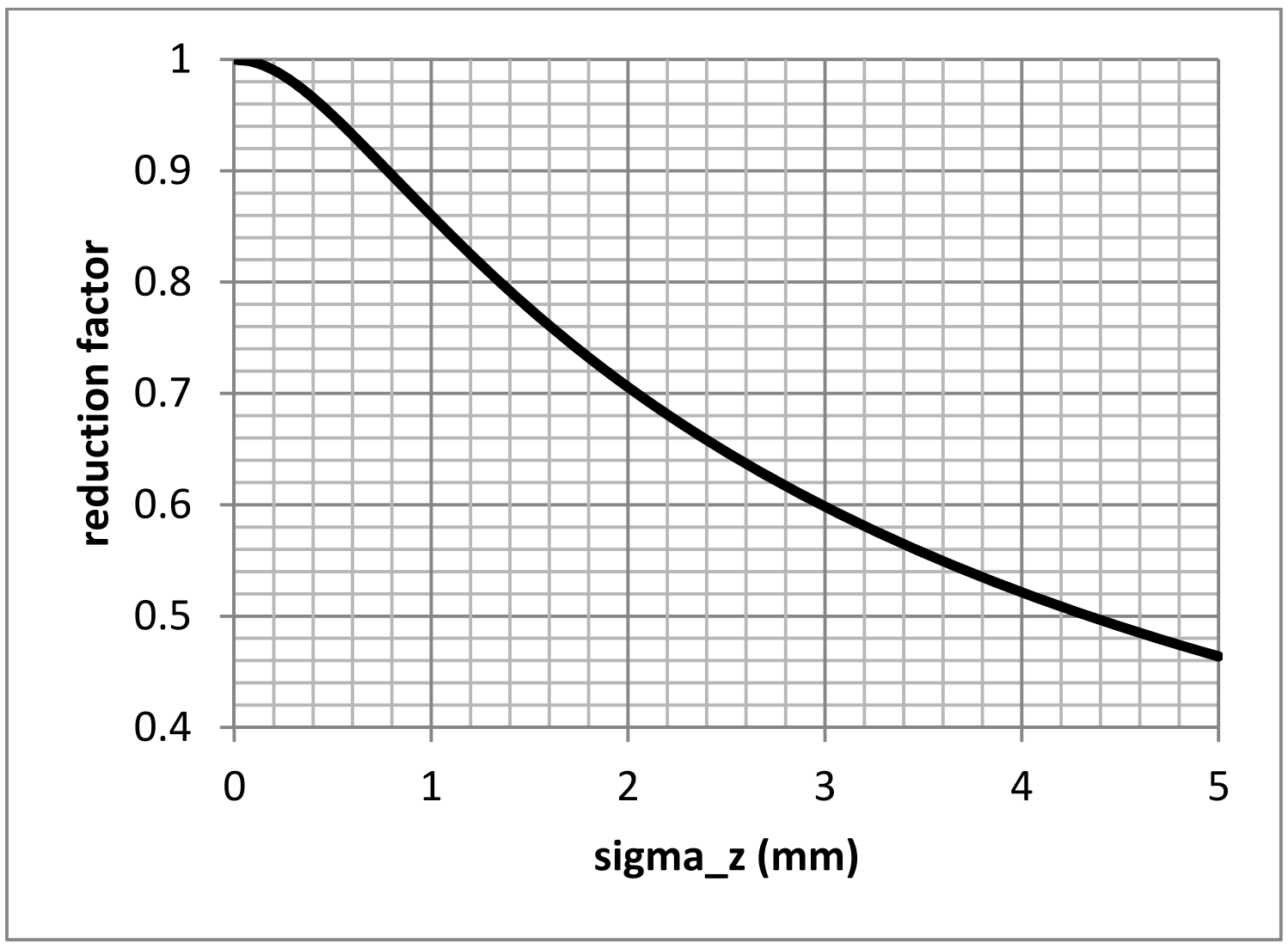

**Figure 3:** Variation of the hourglass factor as the longitudinal beam size changes from 0 to 5mm, while the vertical $\boldsymbol{\beta}^*$ is 1mm

### 3.2. The beam-beam parameter

The strength of the (non-linear) lens effects when one beam crosses the other at the interaction point is measured by the beam-beam parameter $\xi$ which in the vertical plane is defined by eqn. (3). For low beam-beam parameter values, $\xi$ is equal to the observed tune shift. There exists a maximum value for $\xi$ which essentially is a function of the amount of damping present between interaction points, the damping decrement $\lambda_d$:

$$\xi_y{}^{max} = f(\lambda_d) \tag{9}$$

Where

$$\lambda_d = \frac{1}{f_{rev}\,\tau\,n_{IP}} \tag{10}$$

Or more conveniently

$$\lambda_d = \left(\frac{U_0}{E}\right)\frac{1}{n_{IP}} \tag{11}$$

And in terms of beam energy and bending radius

$$\lambda_d \propto \frac{{E_0}^3}{\rho\ n_{IP}} \tag{12}$$

There is considerable debate as to the exact form of the function $f(\lambda_d)$. Reference [2] find that the best fit to the LEP data is ${\xi_y}^{max} \propto {\lambda_d}^{0.4}$.

Figure 4 shows the relative luminosity of a circular collider, compared to a circular collider of bending radius of 10 km, as a function of bending radius for different functions for ${\xi_y}^{max}$. A 5 km radius accelerator, for instance, would deliver between 0.5 and 0.66 of the luminosity of a 10 km accelerator, depending on if ${\xi_y}^{max}$ is independent of the damping decrement or if it is of the form ${\xi_y}^{max} \propto {\lambda_d}^{0.4}$.

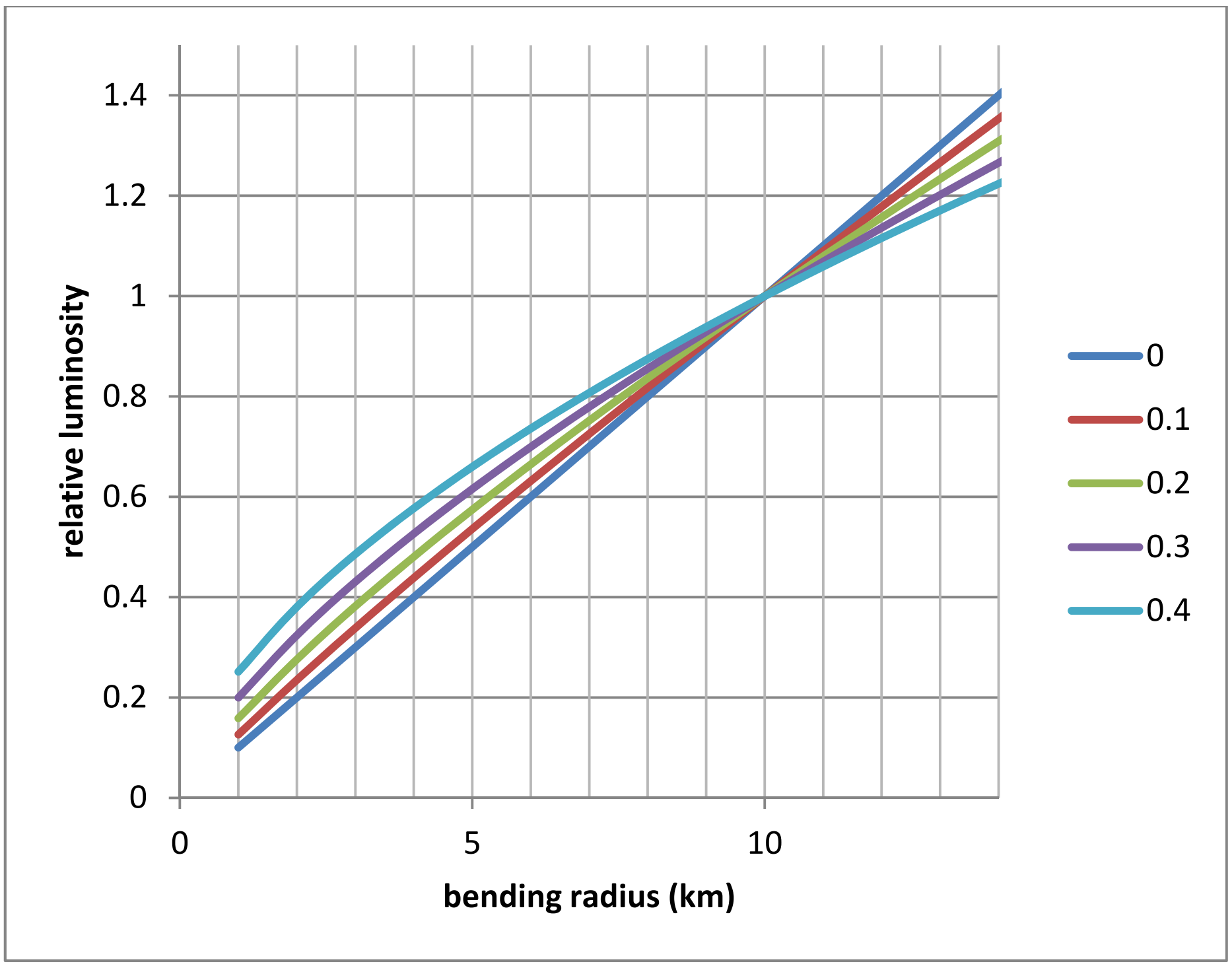

**Figure 4:** Relative luminosity of circular accelerators versus their bending radii (compared to an accelerator with bending radius of 10km), for different assumptions of the relationship of $\boldsymbol{\xi_y}^{\boldsymbol{max}}$ and the damping decrement. We show here curves obtained for different values of the exponent *p* assuming the relationship between the maximum beam-beam parameter and the damping decrement to be of the form $\boldsymbol{\xi_y}^{\boldsymbol{max}} \propto \boldsymbol{\lambda_d}^{\boldsymbol{p}}$

## 4. Beamstrahlung

The relevance of the beamstrahlung process in hindering the performance of circular colliders has recently been pointed out by V. Telnov [4].

Beamstrahlung is the synchrotron radiation emitted by an incoming electron in the collective electromagnetic field of the opposite bunch at an interaction point. The main effect at circular colliders at high energy is a (single) hard photon exchange taking the electron out of the momentum acceptance of the machine. If too many electrons are lost, the beam lifetime is affected.

At low energies in circular colliders the main effect of beamstrahlung is an increase of equilibrium energy spread and bunch length [5].

## 4.1. Beamstrahlung lifetime

The energy spectrum of emitted photons during a collision of two intense bunches (this is the usual bremsstrahlung formula) is characterized by a critical energy

$$E_c = \frac{\hbar 3{\gamma_0}^3 c}{2\rho} \quad (13)$$

where ρ is the radius of curvature of the affected electron which depends on the field he sees

$$\rho = \frac{\gamma_0 mc^2}{eB} \quad (14)$$

And in the case of head-on collisions the maximum field for flat beams can be approximated by

$$B_{max} = \frac{2eN_b}{\sigma_x \sigma_z} \quad (15)$$

So $\rho$ for the maximum field becomes

$$\rho = \frac{\gamma_0 \sigma_x \sigma_z}{2r_e N_b} \quad (16)$$

And the critical energy is equal to

$$E_c = E_0 \frac{3{r_e}^2 \gamma_0 N_b}{\alpha \sigma_x \sigma_z} \quad (17)$$

where $r_e$ is the classical radius of the electron $r_e = \frac{e^2}{mc^2}$ , $\alpha$ the fine structure constant $\alpha = \frac{e^2}{\hbar c}$ , and $\gamma_0 = E_0/mc^2$. This is the critical energy for the maximum field, it would be smaller for a lower field.

Telnov uses the approximation that 10% of electrons see the maximum field, and 90% see no field at all. This is an approximation that needs to be verified with simulation. We call this fraction $f_{max.field}$

$$f_{max.field} = 0.1 \quad (18)$$

Electrons are lost if they emit a gamma with energy larger than the momentum acceptance, η: $E_\gamma \geq \eta E_0$

We define

$$u = \eta \frac{E_0}{E_c} \quad (19)$$

Or otherwise

$$u = \frac{\alpha}{3\gamma {r_e}^2} \eta \frac{\sigma_x \sigma_z}{N_b} \quad (20)$$

The number of photons with $E_\gamma \geq \eta E_0$ is given by:

$$n_\gamma = \frac{f_{max.field}\alpha^2\eta l}{\sqrt{6\pi} r_e \gamma u^{3/2}} e^{-u} \quad (21)$$

Where $l$ is the collision length, approximated by $l = \frac{\sigma_z}{2}$.

The interaction frequency (revolution frequency times number of interaction points) is given by

$$f_{interaction} = \frac{n_{IP} c}{2\pi R} \quad (22)$$

Where R is the geometric radius of the machine (and not the bending radius). Therefore, the beam lifetime due to beamstrahlung (BS) is $\tau_{BS} = 1/n_\gamma f_{interaction}$

$$\tau_{BS} = \frac{4\pi R}{n_{IP} c} \frac{\sqrt{6\pi} r_e \gamma u^{3/2}}{f_{max.field}\alpha^2 \eta \sigma_z} e^u \quad (23)$$

or

$$\tau_{BS} = \frac{4}{3}\frac{\pi R}{n_{IP} c}\sqrt{\frac{2\pi\eta}{\alpha\gamma}}\frac{1}{\sigma_z {r_e}^2}\left(\frac{\sigma_x \sigma_z}{N_b}\right)^{3/2}\frac{1}{f_{max.field}} e^u \quad (24)$$

A. Bogomyagkov et al. [6] arrive at a slightly different formula

$$\tau_{BS} = \frac{8}{3}\frac{\pi R}{n_{IP} c}\sqrt{\frac{2\eta}{\alpha\gamma}}\frac{1}{\sigma_z {r_e}^2}\left(\frac{\sigma_x \sigma_z}{\sqrt{2} N_b}\right)^{3/2} e^u \quad (25)$$

Where in this case

$$u = \frac{\sqrt{2}\alpha}{3\gamma {r_e}^2}\eta\frac{\sigma_x \sigma_z}{N_b} \quad (26)$$

It should be noted that the problem of BS lifetime becomes important at high energies. This is because for a specific ring, power consumption, emittances and $\xi$, the number of particles per bunch scales with gamma as can be seen by rearranging eqn. (3):

$$N_b = \xi_y \frac{2\pi\gamma\sigma_x\sigma_y}{r_e \beta_y^*} \quad (27)$$

And, therefore, $u$ scales with $\gamma^{-2}$. This produces a steep drop in lifetime with increased energy.

The beam BS lifetimes depend on the specific implementation. Essentially, they depend on $(\eta\sigma_x\sigma_z)$ which should be maximized. Also, the minimum tolerable beam lifetime would depend on accelerator specifics, namely on the period of top-up injection. It should be noted here that beam lifetimes due to unavoidable Bhabha interactions for such high-luminosity implementations are O($10^3$ sec), therefore a top-up scheme with a period shorter than this is mandatory. Implementation [1] has a top-up period of O(10s).

The BS lifetimes calculated with both formulas can be seen in Figure 5. Here we have fixed the momentum acceptance at 2% and the other accelerator parameters to the TLEP parameters for 175GeV running [1] with the exception of the vertical beam-beam parameter where we use the two values, $\xi_y = 0.056$ and $\xi_y = 0.09$ , corresponding to the 175GeV and 120GeV TLEP design. For the specific TLEP design, the fill up period is of O(10s), meaning that beam lifetimes above 100 s are tolerable, and about 1000 s comfortable. The difference between the two formulas in the region of most interest (above 170GeV, where BS lifetimes become short), is small (maximum factor of 3).

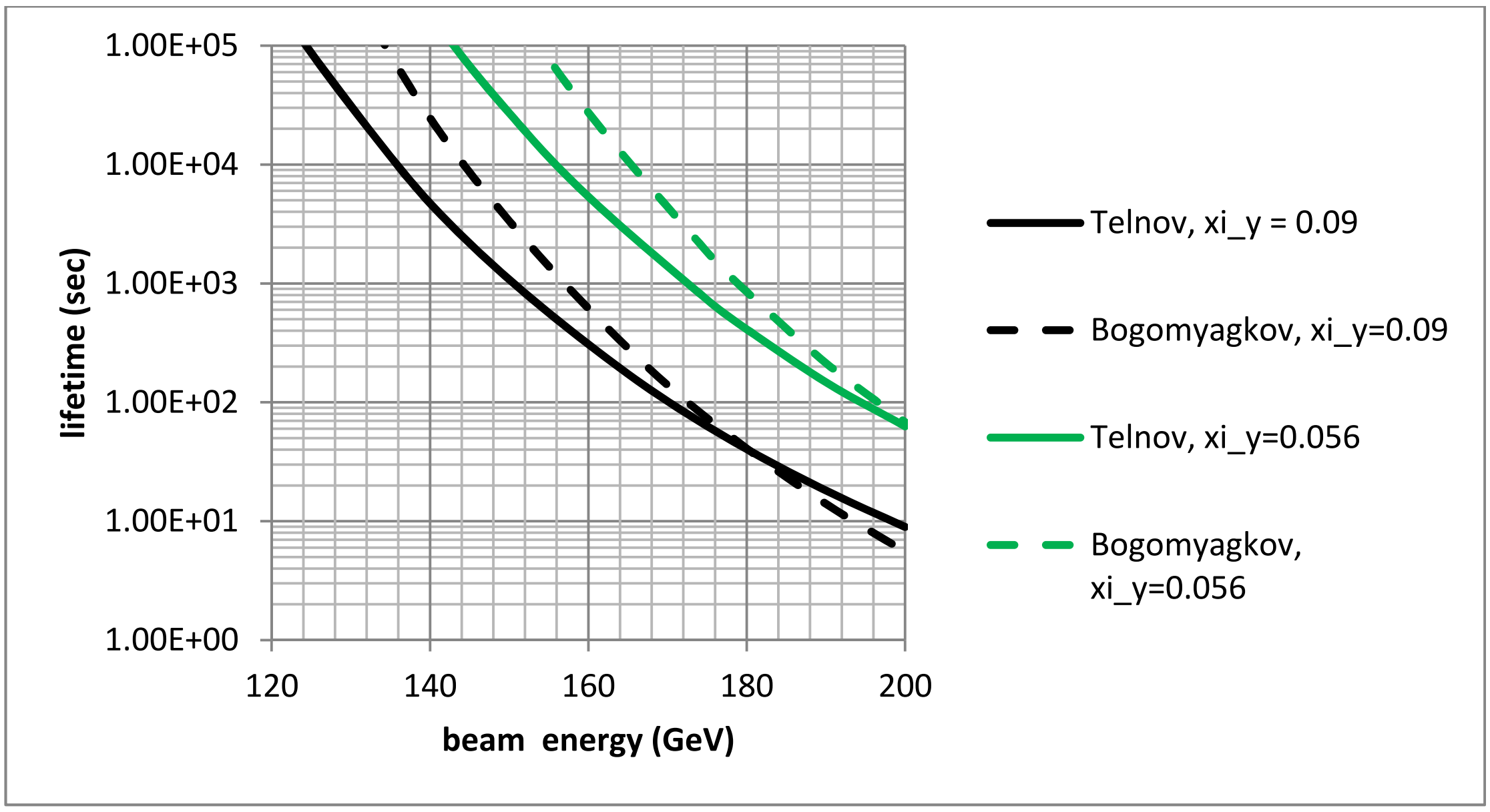

**Figure 5:** Beam lifetimes due to beamstrahlung following the specific implementation in [1] and a momentum acceptance of 2% for two different beamstrahlung formulas and two vertical beam parameters.

It should be stressed that both formulas contain approximations, so simulation would be an important tool in verifying any design.

## 4.2. Beamstrahlung - induced energy spread and bunch length

Another important effect of beamstrahlung, more relevant at low energies, is the increase of the energy spread and therefore bunch length. Since the beamstrahlung effect depends on bunch length (the larger the length the smaller the effect), an equilibrium bunch length is reached where the BS effect is just big enough to sustain it. The BS energy spread and bunch length contributions need to be added to other contributions, notable the contribution due to synchrotron radiation.

For the formulation we use the approach and approximations in [1].

The strength of the beamstrahlung is characterized by a parameter Ψ that can be expressed as

$$\Psi \approx \frac{5}{6}\frac{r_e^2 \gamma N}{\alpha \sigma_z (\sigma_x + \sigma_y)} \tag{28}$$

The average number of photons emitted per collision is roughly given by

$$n_\gamma \approx 2.54 \left[ \frac{\alpha^2 \sigma_z}{r_e \gamma} \frac{\Psi}{(1 + \Psi^{2/3})^{1/2}} \right] \tag{29}$$

and the average relative energy loss by

$$\delta_B \approx 1.24 \left[ \frac{\alpha^2 \sigma_z}{r_e \gamma} \frac{\Psi^2}{\left(1 + (1.5\Psi)^{\frac{2}{3}}\right)^2} \right] \tag{30}$$

The additional relative energy spread due to beamstrahlung can now be estimated to be

$$\sigma_{E,BS} \approx \frac{1}{2} \sqrt{\frac{\tau_E n_{IP}}{T_0}}\, \delta_B \left[ 0.333 + \frac{4.583}{n_\gamma} \right]^{\frac{1}{2}} \tag{31}$$

Where $\frac{\tau_E}{T_0}$ is the longitudinal damping time expressed as number of turns. This additional relative energy spread can now be added in quadrature to the energy spread due to other factors (primarily the SR energy spread) to obtain the total relative energy spread. The self-consistent full energy spread can then be obtained iteratively, as the BS energy spread depends on the bunch length, itself a function of energy spread.

## 5. Luminosity limitations – the two regimes

We have seen that there are essentially two limitations on the maximum achievable luminosity of a circular collider, given its dimensions and power consumption. At low energies, performance is dominated by the beam-beam limit, whereas at high energies by the Beamstrahlung lifetime. To illustrate at what beam energy the switchover from a beam-beam dominated machine to a beamstrahlung dominated one happens, we have chosen to use beam-beam parameter space (Figure 6). The beamstrahlung curve is obtained by fixing the BS lifetime to 300 s and noting at what vertical beam-beam parameter this happens. The machine bending radius is 11 kms, the vertical beta* 1mm, the vertical emittance 2pm and the momentum acceptance 2%. The bunch length is the equilibrium bunch length taking into account the beamstrahlung and the synchrotron radiation contributions. The synchrotron radiation contribution to the bunch length is not constant but computed from a fixed momentum compaction factor $a_c = 5 \times 10^{-6}$, synchrotron tune $Q_s = 0.1$ and longitudinal damping partition number $J_s = 2$ , that are taken from the 175 GeV parameter set of reference [1].

As we go higher in energy, to keep the same lifetime we need to increase the number of bunches (or decrease the number of electrons per bunch). We use the formula of reference [6] for the BS lifetime computation. Note that in this representation ( in vertical beam-beam parameter space) the resulting curve does not depend on the horizontal emittance.

As can be seen from the figure, with the parameters stated above the switchover happens at around 145-170GeV depending on the momentum acceptance of the machine (in this case 1.5 or 2%), above which the machine cannot run at its theoretical maximum beam-beam parameter. At 175GeV the performance of the machine is reduced by about 25% to 50% (for momentum acceptance values of 2% or 1.5% respectively), compared to a purely beam-beam dominated machine with these parameters.

**Table 1:** The cross-over beam energy between a beam-beam dominated machine and a beamstrahlung dominated one using the assumptions in the text (beam lifetimes of 300 seconds, beamstrahlung formula from [6], momentum acceptance of 2% ).

| **Machine circumference (km)** | **Bending radius (km)** | **Cross-over beam energy (GeV)** |
|---|---|---|
| 27 | 3.0 | 126 |
| 50 | 5.5 | 144 |
| 70 | 7.7 | 155 |
| 80 | 8.8 | 160 |
| 100 | 11.0 | 169 |

For smaller machines, the maximum beam-beam parameter curve moves up due to larger damping, and for example a machine with a bending radius of 3.1kms (LEP3) would become beamstrahlung dominated above 126GeV (everything else being equal). The cross-over for different size rings are shown in Table 1. The assumptions used are a vertical emittance of 2pm, a fixed contribution to the bunch length due to synchrotron radiation of 1.2mm, vertical beta* of 1mm, and momentum acceptance of 2%. It is interesting to note that all machines are beamstrahlung dominated at the top threshold (175 GeV) and none is beamstrahlung dominated at the Higgs running (120 GeV).

To push the cross-over to higher energies, we need to reduce the vertical emittance further than the 2 pm assumed here, or have higher momentum acceptance than the currently assumed 2%, or increase the longitudinal bunch length due to SR. Performing the latter decreases the luminosity due to the hourglass effect, but eases the BS problem, ultimately allowing for higher luminosity that not only compensates the loss due to the hourglass effect, but allows a net gain. Changing the emittance and therefore the beam size horizontally does not change the beam-beam parameter or BS lifetime. Simply the number of electrons per bunch compensate for this change in horizontal beam size.

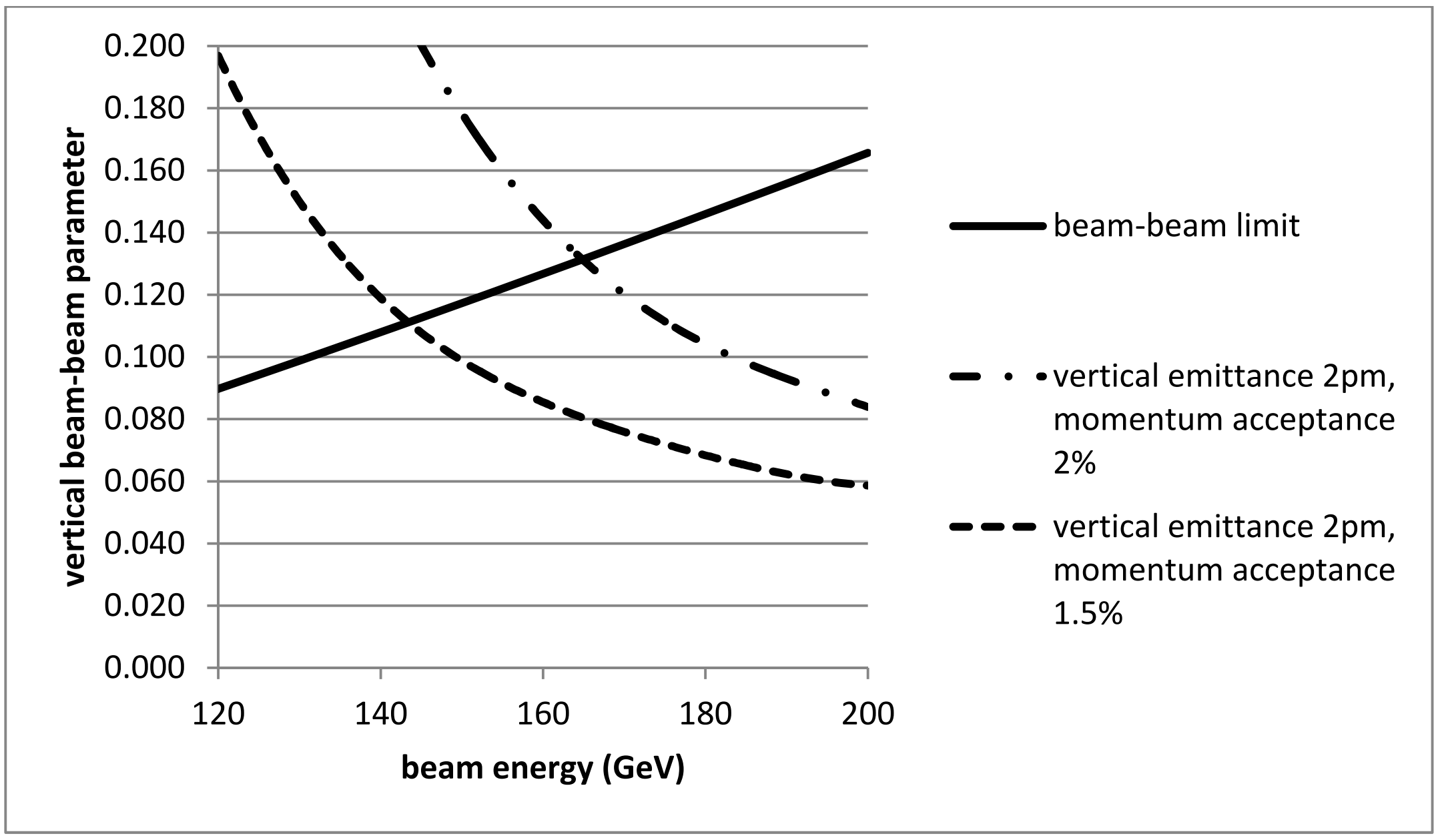

**Figure 6:** Cross over between beam-beam dominated and beamstrahlung dominated machine for a circular collider with circumference 100 kms and the rest of the parameters as in the parameter table of [1] for a machine running at 175 GeV: Vertical emittance of 2 pm, vertical beta* of 1mm, and momentum acceptance of 2%. The longitudinal beam size due to synchrotron radiation varies with energy using $\boldsymbol{a_c = 5 \times 10^{-6}}$, $\boldsymbol{Q_s}$= 0.1 and $\boldsymbol{J_s}$ = 2. The overall longitudinal bunch length is the equilibrium one. In this case the crossover is 144GeV for momentum acceptance of 1.5% and 169GeV for momentum acceptance of 2%. The solid curve shows the energy dependence of the vertical beam-beam parameter assuming a $\boldsymbol{\lambda_d}^{\mathbf{0.4}}$ dependence and a value of 0.09 at 120GeV as in [1].

## 6. Conclusions

We have presented the main limitations of a circular $e^+e^-$ collider, applicable to the suite of circular colliders being discussed following the Higgs boson discovery. The limitations concern energy reach, maximum achievable luminosity and beam lifetimes due to beamstrahlung. The luminosity of such a collider is proportional to the SR power consumed in the arcs, and rises with the collider bending radius. The beamstrahlung lifetime depends on the specific implementation (momentum acceptance, vertical emittance, vertical beta*, ring size, top-up frequency) . For the implementation of FCC-ee [1], the machine becomes beamstrahlung dominated above 150GeV beam energy.

Collisions at an angle and more exotic schemes are not covered here and merit a separate publication.

## 7. Acknowledgements

We would like to thank the early pioneers of modern circular Higgs factories for their inspiration and encouragement (Blondel, Zimmermann, Ellis, Janot, Aleksan). We would also like to thank Jorg Wenninger, Frank Zimmermann, (complete as necessary) for valuable comments.

# 8. References

[1] Wenninger, J. et al., "Future Circular Collider Study Lepton Collider Parameters," *CERN EDMS no. 1346082,* 2014.

[2] R. Assmann and K. Cornellis, "The beam-beam limit in the presence of strong synchrotron radiation damping," *CERN-SL-2000_046 OP.*

[3] A. W. Chao and M. Tigner, Handbook of Accelerator Physics and Engineering, Singapore: World Scientific, 1999.

[4] V. Telnov, "Restriction on the energy and luminosity of e+e- storage rings due to beamstrahlung," *Phys. Rev. Letters 110, 114801 (2013) arXiv:1203.6563.*

[5] K. Yokoya, P. Chen, "Beam-Beam Phenomena in Linear Colliders," *Lect.Notes Phys.,* vol. 400, pp. 415-445, 1992.

[6] A. Bogomyagkov et al, "Beam-beam effects investigation and parameters optimization for a circular e+e- collider TLEP to study the Higgs boson," *arXiv:1311.1580v1.*